\documentclass{aa}

\usepackage[varg]{txfonts}
\usepackage[colorlinks=true, linkcolor=blue, citecolor=blue, urlcolor=gray]{hyperref}
\usepackage{xcolor}
\usepackage{listings}

\newcommand{\txd}{{\text{d}}}
\newcommand{\txe}{{\text{e}}}
\newcommand{\bfx}{{\boldsymbol{x}}}
\newcommand{\bfn}{{\boldsymbol{n}}}

\begin{document}

\title{Monte Carlo radiative transfer with explicit absorption \\
to simulate absorption, scattering, and stimulated emission}

\titlerunning{MCRT with explicit absorption}

\author{Maarten Baes, Peter Camps, and Kosei Matsumoto}

\institute{Sterrenkundig Observatorium, Universiteit Gent, Krijgslaan 281 S9, B-9000 Gent, Belgium}

\date{Received 19 July 2022 / Accepted 11 August 2022}

\abstract{The Monte Carlo method is probably the most widely used approach to solve the radiative transfer problem, especially in a general 3D geometry. The physical processes of emission, absorption, and scattering are easily incorporated in the Monte Carlo framework. Net stimulated emission, or absorption with a negative cross section, does not fit this method, however.}%
{We explore alterations to the standard photon packet life cycle in Monte Carlo radiative transfer that allow the treatment of net stimulated emission without loss of generality or efficiency.}%
{We present the explicit absorption technique that allows net stimulated emission to be handled efficiently. It uses the scattering rather than the extinction optical depth along a photon packet's path to randomly select the next interaction location, and offers a separate, deterministic treatment of absorption. We implemented the technique in a special-purpose Monte Carlo code for a two-stream 1D radiative transfer problem and in the fully featured 3D code SKIRT, and we studied its overall performance using quantitative statistical tests.}%
{Our special-purpose code is capable of recovering the analytical solutions to the two-stream problem in all regimes, including the one of strong net stimulated emission. The implementation in SKIRT is straightforward, as the explicit absorption technique easily combines with the variance reduction and acceleration techniques already incorporated. In general, explicit absorption tends to improve the efficiency of the Monte Carlo routine in the regime of net absorption.}%
{Explicit absorption allows the treatment of net stimulated emission in Monte Carlo radiative transfer, it interfaces smoothly with other variance reduction and acceleration techniques, and it tends to improve the efficiency of the simulations in the net absorption regime. We recommend to always include this new technique in Monte Carlo radiative transfer.}

\keywords{Radiative transfer -- Methods: numerical}

\maketitle

%%%%%%%%%%%%%%%%%%%%%%%%%%%%%%%%%%%%%%%%%%%%%%%%%%%%%%%%%%%%%%%

\section{Introduction}

Radiative transfer is the general process that describes the interaction and transfer of energy when electromagnetic radiation passes through a medium. In astronomy, where we need to extract virtually all of the information on the Universe from radiation, radiative transfer is relevant in every field, from exoplanets to cosmology. General reviews on astrophysical radiative transfer include the seminal works by \citet{1960ratr.book.....C} and \citet{1979rpa..book.....R}.

The Monte Carlo method is probably the most widely used approach to deal with radiative transfer problems \citep[e.g.,][]{2011BASI...39..101W, 2013ARA&A..51...63S, 2019LRCA....5....1N}. The essence of the Monte Carlo radiative transfer (MCRT) technique is to represent the radiation field as the flow of a huge number of individual photon packets. The life cycle of each photon packet is followed individually through the medium, whereby each step in this cycle is determined in a probabilistic way by generating random numbers from the appropriate probability density function (PDF). The radiation field is constructed from the statistical analysis of the trajectory of all the different photon packets. Important advantages of the MCRT method are its flexibility and its inherent 3D nature, which makes the method particularly attractive for complex geometries. 

The physical processes of emission, absorption, and scattering are easily incorporated in the MCRT framework. These processes are the most important ones for continuum radiative transfer, for example dust radiative transfer \citep{2013ARA&A..51...63S}. An additional process that needs to be considered when performing non-LTE line radiative transfer calculations is stimulated emission. By this process, a photon of a specific wavelength interacts with an atom or molecule in an excited state, causing it to drop to a lower energy level and liberating the internal energy in the form of a new photon. In many circumstances, stimulated emission can easily be incorporated in the MCRT framework `under the hood', as absorption and stimulated emission are reverse processes proceeding at different rates, and one can combine them into a single process \citep[e.g.,][]{2000A&A...362..697H, 2007A&A...468..627V, 2010A&A...523A..25B}. In most media, the absorption rate exceeds the stimulated emission rate because there are more electrons in the lower energy states than in the higher energy states. The combination of both processes then results in a net absorption cross section. 

There are, however, situations where this assumption is not valid. When a population inversion is present, the stimulated emission rate exceeds the absorption rate, resulting in a net stimulated emission cross section, or equivalently, a formally negative absorption cross section. The result is that the intensity of the radiative field increases exponentially when it passes through the medium, a physical phenomenon known as optical amplification. In astrophysics, optical amplification is mainly important for masers. Traditional masers are produced by rotational or vibrational transitions in molecules such as OH, H$_2$O or SiO, when the density is considerably higher than in the diffuse ISM and a luminous source acts as a pumping mechanism to generate a population inversion \citep{1981ARA&A..19..231R, 1992ARA&A..30...75E, 2005ARA&A..43..625L}. Another class consists of hydrogen recombination line masers, where the recombination lines are amplified by stimulated emission in the Rayleigh-Jeans limit, even at low optical depths \citep{1978ApJS...37..459K, 1989A&A...215L..13M, 2018ApJ...867L...6B}.

The presence of net stimulated emission, or absorption with a negative cross section, does not fit the standard MCRT method. Indeed, one of the core ingredients in MCRT is the conversion of a randomly generated optical depth to a physical path length \citep{2003A&A...399..703N, 2003MNRAS.343.1081B, 2019A&A...630A..61B, 2011BASI...39..101W, 2013ARA&A..51...63S}. If the cross section along a portion of the path is negative, the optical depth is no longer a monotonic function of the distance along the path, rendering the conversion impossible -- or at least ambiguous -- and this fundamental MCRT ingredient breaks down. One could attempt to incorporate net stimulated emission in MCRT using an iterative approach. First perform an MCRT phase that ignores stimulated emission. When this is finished, launch additional photon packets corresponding to stimulated emission in a second phase, and iterate this phase until convergence is reached. This approach has substantial disadvantages. The second iterative phase adds a significant layer of complexity and convergence might be slow because of the expected exponential growth of the radiation field intensity. Moreover, this approach requires storing the radiation field including directional information at every location, which is extremely expensive for 3D simulations. 

In this paper, we present a technique, which we call explicit absorption, that allows us efficiently handle net stimulated emission, or equivalently, absorption with negative cross sections, within a single MCRT phase. The text is organised as follows. We first describe the method in Sect.~\ref{sec:Theory} and then verify its correctness using a straightforward 1D problem that can be solved analytically in Sect.~\ref{sec:Application1D}. Subsequently, we discuss the implementation of the explicit absorption technique in our 3D MCRT code SKIRT \citep{2020A&C....3100381C} in Sect.~\ref{sec:ImplementationSKIRT}, verify its operation using an idealised 3D configuration and published benchmark problems in Sect.~\ref{sec:Application3D}, and study its efficiency compared to the regular MCRT method in Sect.~\ref{sec:Efficiency}. Finally, we summarise and conclude in Sect.~\ref{sec:Conclusions}.

%%%%%%%%%%%%%%%%%%%%%%%%%%%%%%%%%%%%%%%%%%%%%%%%%%%%%%%%%%%%%%%

\section{Monte Carlo radiative transfer with stimulated emission}
\label{sec:Theory}

\subsection{The radiative transfer equation}

The standard time-independent monochromatic radiative transfer equation can be written as
\begin{multline}
\frac{\txd I}{\txd s}(\bfx,\bfn)
=
j(\bfx,\bfn) - n(\bfx)\,C_{\text{abs}}(\bfx)\,I(\bfx,\bfn) - n(\bfx)\,C_{\text{sca}}(\bfx)\,I(\bfx,\bfn) 
\\
+ n(\bfx)\,C_{\text{sca}}(\bfx) \int I(\bfx,\bfn')\,\Phi(\bfn,\bfn')\,\txd\Omega'.
\label{RTE}
\end{multline}
In this expression, $\bfx$ and $\bfn$ represent the position and the propagation direction of the radiation, respectively. $I$ represents the specific intensity of the radiation, $j$ the emissivity, $n$ the number density of the medium, $C_{\text{abs}}$ and $C_{\text{sca}}$ the absorption and scattering cross sections, and $\Phi$ the scattering phase function. In most cases, the second and third terms on the right-hand side of this equation are combined in a single extinction term, with the extinction cross section
\begin{equation}
C_{\text{ext}} = C_{\text{abs}} + C_{\text{sca}}
\end{equation}
simply the sum of the absorption and scattering cross sections. We prefer to write them explicitly to stress the two processes at work. In the case of stimulated emission, the absorption cross section is the sum of the positive contribution of actual absorption and the negative contribution of stimulated emission. When stimulated emission dominates over absorption, we find that $C_{\text{abs}}<0$.

For the sake of simplicity, we assume in this paper that the absorption and scattering cross sections are independent of the intensity of the radiation field. In real astrophysical situations, the level populations of the atoms and molecules in the medium, and thus the absorption and scattering coefficients, do depend on the radiation field strength. In particular, a population inversion will be reduced when the radiation field intensity is sufficiently high, ultimately leading to a saturation of the radiation field \citep[e.g.,][]{1981ARA&A..19..231R, 1990ApJ...363..628E, 1990ApJ...363..638E}. The simplifying assumption we make comes down to assuming that we remain in the unsaturated regime. 

\subsection{Basic Monte Carlo radiative transfer}
\label{BasicMC.sec}

The simplest MCRT calculation considers radiation at a single wavelength. Throughout the calculation, the state of each individual photon packet is characterised by its weight $W$ (equivalent to the number of photons within the packet), its position $\bfx$, and its propagation direction $\bfn$. Once the photon packet is launched into the medium, its life cycle is determined by a loop. This loop consists of three steps, during which the properties of the packet are updated.

The first step in the photon packet life cycle (PLC) consists of randomly determining whether the photon packet will interact with the medium, and if so, where this interaction will take place. The PDF that describes the free path length $s$ before an interaction is most conveniently described in optical depth space, where it has an exponential distribution. In the light of the remainder of this section, it is useful to revisit the reasoning behind this. For a photon packet moving through an attenuating medium with density $n(s)$ and extinction cross section $C_{\text{ext}}(s)$, the radiative transfer equation reads
\begin{equation}
\frac{\txd I}{\txd s}(s) = -n(s)\,C_{\text{ext}}(s)\,I(s),
\end{equation}
where we note that no additional emission processes are relevant for this particular photon packet. The solution of this simple ordinary differential equation is readily found,
\begin{equation}
I(s) = I(0)\,\txe^{-\tau_{\text{ext}}(s)},
\label{expdecline}
\end{equation}
with the monotonically increasing optical depth scale $\tau_{\text{ext}}(s)$ defined as 
\begin{equation}
\tau_{\text{ext}}(s) = \int_0^s n(s')\,C_{\text{ext}}(s')\,\txd s'.
\label{tauexts}
\end{equation}
The exponential decline of the intensity in reality means that individual photons are continuously disappearing from the photon packet due to interactions with the medium as the photon packet propagates. Ideally, we should model all of these interactions individually. The essence of the Monte Carlo method consists of replacing this continuous behaviour by stochastically sampling a single interaction location for the entire photon packet. The challenge is to find the appropriate PDF $p(s)$ from which we should sample a physical path length before the next interaction. This PDF can be found by noting that the probability that a photon does not interact along the path between $0$ and $s$ is equal to $I(s)/I(0)$. The cumulative distribution $P(s)$ corresponding to the desired PDF $p(s)$ is thus
\begin{equation}
P(s) = 1-\frac{I(s)}{I(0)} = 1-\txe^{-\tau_{\text{ext}}(s)},
\label{Ps}
\end{equation}
or in optical depth space,
\begin{equation}
P(\tau_{\text{ext}}) = 1-\txe^{-\tau_{\text{ext}}}.
\end{equation}
The PDF is found by differentiation, and turns out to be a simple exponential distribution,
\begin{equation}
p(\tau_{\text{ext}}) = \txe^{-\tau_{\text{ext}}}.
\end{equation}
Generating a random $\tau_{\text{ext}}$ from an exponential distribution is straightforward. If this random optical depth exceeds the total optical depth along the path, 
\begin{equation}
\tau_{\text{ext, path}} = \int_0^\infty n(s)\,C_{\text{ext}}(s)\,\txd s,
\end{equation}
the photon packet survives the journey along the path and exits the medium. If, on the other hand, the random optical depth is smaller than the total optical depth, we can calculate the physical path length $s$ to the next interaction location by inverting the relation~(\ref{tauexts}). For complex 3D models, this inversion is often the most time-consuming part of the entire PLC. It is usually accomplished through the use of discretisation on a spatial grid, combined with efficient grid traversal algorithms \citep[e.g.,][]{2003MNRAS.343.1081B, 2013A&A...560A..35C, 2013A&A...554A..10S, 2014A&A...561A..77S, 2016MNRAS.456..756H}. With the physical path length $s$ determined, the photon packet can be propagated to the next interaction location.

The second step in the PLC, at least if the photon packet has not left the system before the interaction, is the determination of the nature of the interaction at the interaction site. The photon packet can either be absorbed or scattered; the appropriate PDF is hence not a continuous but a discrete one with only two possible values. If the absorption and scattering cross sections at the location $\bfx$ are $C_{\text{abs}}(\bfx)$ and $C_{\text{sca}}(\bfx)$, respectively, the discrete PDF that describes the probabilities for both types is
\begin{subequations}
\begin{gather}
p({\text{abs}}) = \frac{C_{\text{abs}}(\bfx)}{C_{\text{ext}}(\bfx)} = 1 - \omega(\bfx), \label{pabs} \\
p({\text{sca}}) = \frac{C_{\text{sca}}(\bfx)}{C_{\text{ext}}(\bfx)} = \omega(\bfx), \label{psca}
\end{gather}%
\label{pabssca}%
\end{subequations}%
where $\omega$ represents the scattering albedo. Drawing a random interaction type from this discrete PDF is trivial. If the event is an absorption event, the photon packet's life cycle terminates. In the case of a scattering event, its journey continues, and we move to the third step in the cycle.

The third and final step in the PLC is the actual simulation of a scattering event at the location of the interaction. This comes down to generating a new propagation direction for the photon packet, based on the scattering phase function. The details of this step are not very relevant for the discussion in the remainder of this paper.

Each of these three steps are repeated until the photon packet either escapes from the system or is absorbed by the medium. We can then continue with the next photon packet and repeat the entire procedure. This entire process is repeated for all the photon packets until the last one has left the medium. The ultimate goal of a MCRT simulation is usually the determination of the radiation field, either at locations within the model space or corresponding to an external viewpoint. This can be achieved by statistically binning the photon packets according to location, propagation direction and weight. The details of these steps are also less relevant for the remainder of the discussion. 

\subsection{Challenges in the case of net stimulated emission}

The basic MCRT algorithm discussed in the previous section assumed that both the absorption and the scattering cross sections are positive. As discussed in the Introduction, this is not necessarily the case if we consider stimulated emission as a negative contribution to the absorption. In the case that the stimulated emission rate dominates over the absorption rate, we have $C_{\text{abs}}<0$, and we encounter two problems for the MCRT method, depending on the ratio between $C_{\text{abs}}$ and $C_{\text{sca}}$.

If stimulated emission dominates over absorption, but the scattering cross section is sufficiently large such that the total extinction cross section is still positive, we are in what we call the weak net stimulated emission regime. In this regime, the first step in the PLC can proceed as described above. We do have an issue with the second step, the determination of the nature of the interaction, however. Indeed, since $C_{\text{abs}}<0$, we formally have from Eq.~(\ref{pabssca}) that $p({\text{abs}})<0$ and $p({\text{sca}})>1$. This is obviously not a valid discrete PDF from which we can stochastically sample.

If the stimulated emission rate is so large or the scattering cross section so low that $C_{\text{abs}}$ is larger in magnitude than $C_{\text{sca}}$, we enter the strong net stimulated emission regime. In this case the total extinction cross section becomes negative. Concerning the second step in the PLC we have from Eqs.~(\ref{pabssca}) that $p({\text{abs}})>1$ and $p({\text{sca}})<0$, which leads to a similar problem as before. Even more troublesome is that we now encounter a fundamental problem in the first step of the cycle. If the extinction cross section is negative in some regions of the computational domain, we find that the optical depth along each path will no longer be a monotonically increasing function of physical path length, which means that we can no longer interpret Eq.~(\ref{Ps}) as a valid cumulative distribution function. It can even occur that the optical depth is completely negative, which completely breaks the procedure. 

\subsection{The absorption-scattering split}
\label{SplitMC.sec}

The way to deal with the challenges posed by net stimulated emission, or negative absorption cross sections, is to introduce deterministic elements in the Monte Carlo PLC. The introduction of deterministic elements is, next to biasing or variance reduction, usually done to accelerate the algorithm \citep[for an overview, see][Sec.~5.2]{2013ARA&A..51...63S}. As we will show here, they can also be used to extend the application range of MCRT.
A simple example of the introduction of a deterministic element in the PLC is the so-called absorption-scattering split \citep[e.g.,][]{1970A&A.....9...53M, 1977ApJS...35....1W}. The second step in the standard PLC consists of randomly deciding whether the photon packet will absorb or scatter at the interaction site. In reality, there will be both absorption and scattering: a fraction of the photons in the packet will be absorbed, and the remaining fraction will scatter. Rather than stochastically determining the nature of the interaction, we can split the photon packet into two parts: one that is absorbed and one that scatters. We know exactly which fraction should be absorbed and scattered: this is given by $1-\omega$ and $\omega$, respectively. So at the location of the interaction, we replace the original photon packet with weight $W$ that is about to interact by two new photon packets: one with weight $(1-\omega)\,W$ that will be absorbed, and one with weight $\omega\,W$ that will scatter. In practice, since the absorbed energy disappears, we are only concerned with the photon packet that scatters and continues its journey. 

This very simple change to the PLC replaces a stochastic choice of the interaction nature by a deterministic procedure. Since stochasticity always introduces Poisson noise, this straightforward adjustment helps to reduce noise. In other words, by introducing the absorption-scattering split, we need fewer photon packets in our simulation to reach a given level of signal-to-noise compared to a simulation that does not incorporate it. 

The same method can, however, also be used to allow for stimulated emission in the weak net stimulated emission regime. In this regime we have $0<-C_{\text{abs}}<C_{\text{sca}}$, so formally $\omega>1$. If a photon packet with weight $W$ is to interact, we create two new photon packets with weights $(1-\omega)\,W$ and $\omega\,W$, respectively. The former packet has a negative weight, but this does not jeopardise the PLC. As it will be absorbed, this photon package does not travel through the medium and and does not contribute to the radiation field; the negative weight is thus not an issue. In case the absorption rate of a cell in the medium, and subsequently the mean intensity of the radiation field, the level populations, and the emission rate, are calculated by summing the weights of all absorbed photon packets in that cell \citep[e.g.,][]{2001ApJ...551..269G, 2001ApJ...554..615B, 2005NewA...10..523B}, the resulting absorption rate will be negative. A negative absorption rate simply reflects the fact that we have net stimulated emission: since the absorption cross section is negative as well, the standard formulae to determine the mean radiation field intensity can still be used.\footnote{Most modern Monte Carlo codes use the continuous absorption technique \citep{1999A&A...344..282L, 2003A&A...399..703N, 2011ApJS..196...22B} to estimate the mean intensity of the radiation field and the corresponding emission rate. Rather than explicitly summing the weights of all photon packets absorbed in a given cell, this approach uses the weight of all photon packets that pass through this cell. In this case we do not encounter negative values.} The result is that we can continue the PLC with the second photon packet, which has a positive weight that is larger than the original one. In this way we can simulate net stimulated emission, at least in the weak net stimulated emission regime.

\subsection{Explicit absorption}
\label{ExplicitMC.sec}

The absorption-scattering split can be used to deal with stimulated emission in the weak net stimulated emission regime, but not in the strong net stimulated emission regime. In this latter regime we have the more fundamental problem of negative extinction cross sections. We can, however, introduce another deterministic element in the Monte Carlo PLC that can overcome this problem. We call this approach explicit absorption.

The first step in the standard PLC consists of picking a random path length for the next interaction location based on the interpretation of the intensity~(\ref{expdecline}) as a cumulative distribution~(\ref{Ps}). Since the extinction cross section is just the sum of the absorption and scattering cross sections, we can also rewrite Eq.~(\ref{expdecline}) as 
\begin{equation}
I(s) = \left[ I(0)\,\txe^{-\tau_{\text{abs}}(s)}\right] \txe^{-\tau_{\text{sca}}(s)},
\end{equation}
with the absorption and scattering optical depth scales defined as
\begin{gather}
\tau_{\text{abs}}(s) = \int_0^s n(s')\,C_{\text{abs}}(s')\,\txd s',
\label{tauabss}
\\
\tau_{\text{sca}}(s) = \int_0^s n(s')\,C_{\text{sca}}(s')\,\txd s'.
\label{tauscas}
\end{gather}
Based on this equation, we can change the PLC. Rather than simulating the entire extinction stochastically, we split the absorption and scattering contributions. We treat the absorption contribution explicitly and only the scattering contribution stochastically. We implement the explicit absorption by continuously decreasing the intensity, or the weight, of the photon packet as it travels along the path according to
\begin{equation}
W(s) = W(0)\,\txe^{-\tau_{\text{abs}}(s)}.
\label{ws}
\end{equation}
On the other hand, we randomly determine the next scattering location by picking a random scattering optical depth $\tau_{\text{sca}}$ from an exponential PDF, 
\begin{equation}
p(\tau_{\text{sca}}) = \txe^{-\tau_{\text{sca}}},
\end{equation}
and we compare it to the total scattering optical depth along the path,
\begin{equation}
\tau_{\text{sca, path}} = \int_0^\infty n(s)\,C_{\text{sca}}(s)\,\txd s.
\end{equation}
If $\tau_{\text{sca}}>\tau_{\text{sca, path}}$, the photon packet leaves the system. If, on the other hand, $\tau_{\text{sca}}<\tau_{\text{sca, path}}$, the photon packet will scatter along this path. The scattering location is determined by inverting the relation~(\ref{tauscas}) for $s$. In both cases, we should not forget to alter the photon packet's weight according to expression~(\ref{ws}). 

We note that this alteration of the Monte Carlo method reduces the PLC to two steps. Indeed, the second step in the original method, the random determination of the nature of the interaction, disappears. The remaining steps are the determination of and the propagation to the next scattering location, and the simulation of the new direction after the scattering. Since we eliminate a stochastic element from the PLC, this alteration should in principle reduce the noise. On the other hand, it also makes the first step more computationally demanding: we need to determine both the absorption and the scattering optical depth along the path, and take into account the continuous change of the weight of the photon packet. 

The most interesting aspect of this explicit treatment of absorption is that it can also be applied in the case of negative absorption cross sections. The only effect of absorption is the continuous change of the weight of a photon packet as it moves through the medium according to Eq.~(\ref{ws}). Whether $\tau_{\text{abs}}(s)$ is positive or negative, and hence whether the weight decreases or increases as a function of distance along the path, does not really matter. For the determination of the scattering location, on the other hand, it is crucial that the optical depth is a non-decreasing function of distance along the path. Since $C_{\text{sca}}$ is always non-negative, this condition is satisfied. 

%%%%%%%%%%%%%%%%%%%%%%%%%%%%%%%%%%%%%%%%%%%%%%%%%%%%%%%%%%%%%%%

\section{A two-stream radiative transfer problem}
\label{sec:Application1D}

\subsection{Problem definition}

\begin{figure}
\includegraphics[width=\columnwidth]{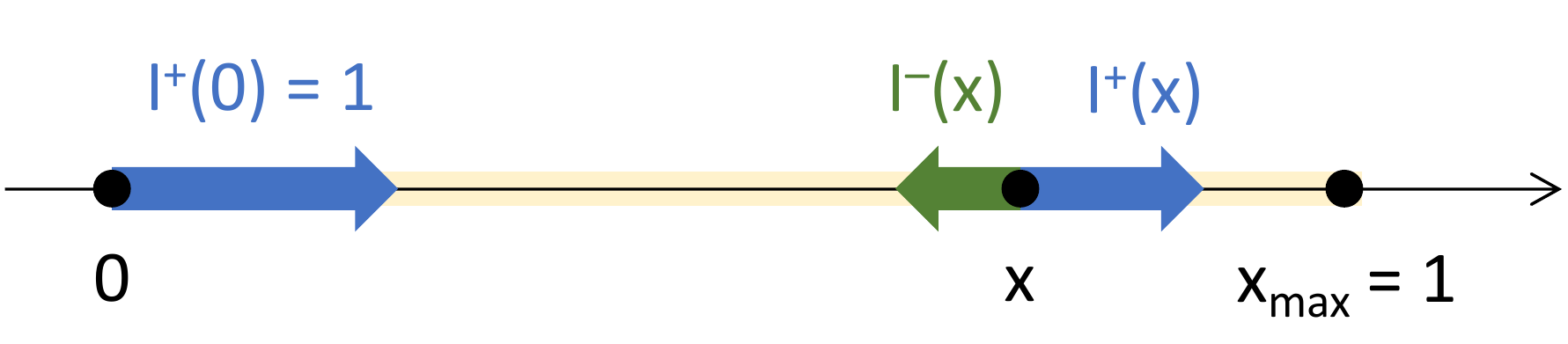}
\caption{Graphical illustration of the two-stream radiative transfer problem considered in Sect.~{\ref{sec:Application1D}}.}
\label{ProblemDefinition.fig}
\end{figure}

We test our new approach using a simple 1D radiative transfer problem for which the solution can be calculated analytically. We consider radiation that propagates through a finite column of matter of length $x_{\text{max}}=1$. We indicate the distance within the column as $x$, and we denote the intensity of the radiation field moving in positive and negative directions as $I^+(x)$ and $I^-(x)$, respectively. A single source of radiation with unit intensity that emits in the positive direction is placed on the left side of the column. The density of matter and the optical properties within the column are uniform, and we arbitrarily set $n=1$. The scattering phase function is such that half of the radiation scatters in the forward direction and the remaining half scatters in the backward direction. The problem is depicted graphically in Fig.~{\ref{ProblemDefinition.fig}}.

With these assumptions, the radiative transfer equation~(\ref{RTE}) can be transformed into a set of two coupled differential equations for $I^+(x)$ and $I^-(x)$,
\begin{equation}
\frac{\txd I^\pm}{\txd x}(x) = 
\mp\left(C_{\text{abs}}+\tfrac12\,C_{\text{sca}}\right) \,I^\pm(x) 
\pm \tfrac12\,C_{\text{sca}}\,I^\mp(x).
\label{RTE1D}
\end{equation}
The boundary conditions are
\begin{gather}
I^+(0) = 1, \label{RTE1D-bound1} \\ 
I^-(x_{\text{max}}) = 0. \label{RTE1D-bound2}
\end{gather}

\subsection{Analytical solution}

\begin{figure}
\includegraphics[width=0.97\columnwidth]{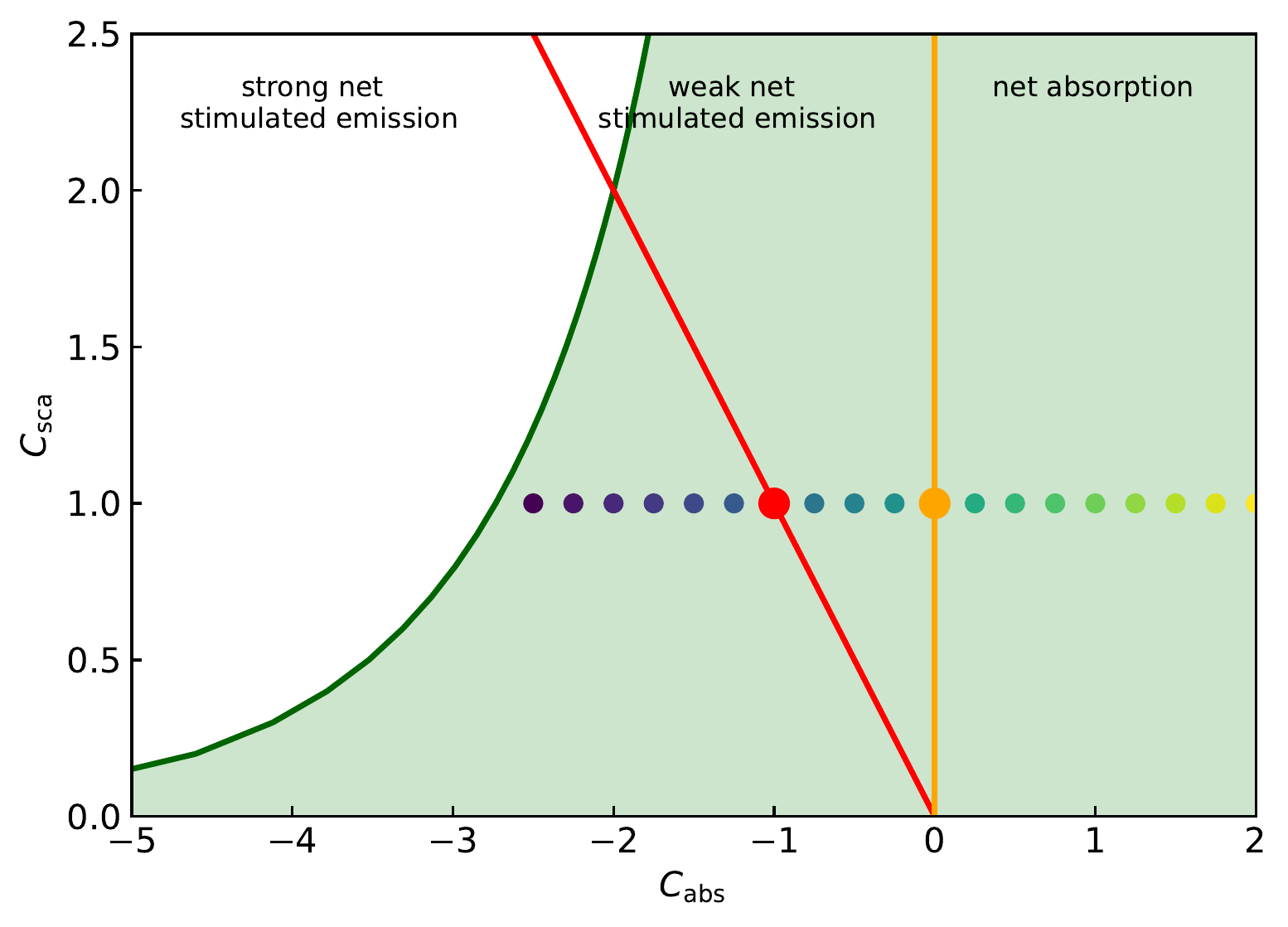}\hspace*{3em}
\caption{$(C_{\text{abs}},C_{\text{sca}})$ parameter space for the 1D radiative transfer model discussed in Sect.~{\ref{simplecase.sec}}. The green shaded area in this plot corresponds to the region with physically viable solutions. The three major regimes (net absorption, weak net stimulated emission, and strong net stimulated emission) are indicated. The orange and red lines indicate the boundaries between these major regimes, and correspond to pure scattering and critical net stimulated emission, respectively. The coloured dots indicate the models shown explicitly in Fig.~{\ref{variableCabs.fig}}. The white area corresponds to non-physical solutions. The solid green line that indicates the boundary between physical and non-physical solutions is given by the Eqs.~(\ref{boundary}) and (\ref{boundary2}).}
\label{variableCabs-paramspace.fig}
\end{figure}

\begin{figure*}
\includegraphics[width=\textwidth]{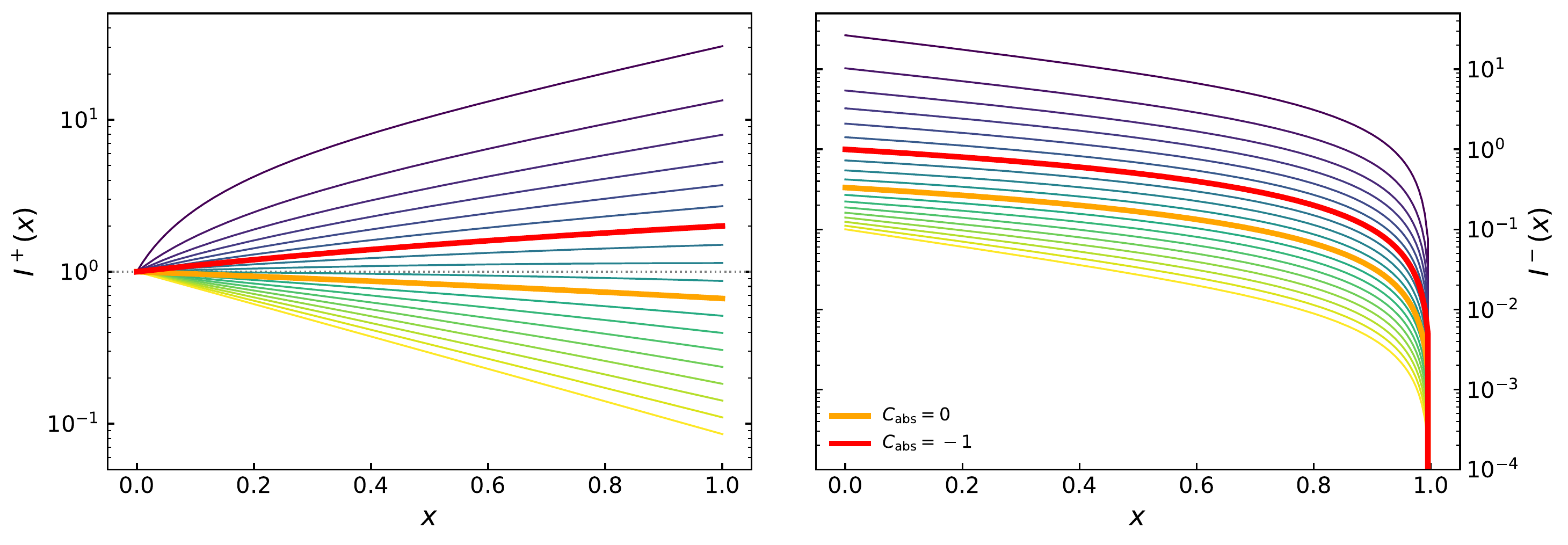}
\caption{Intensity of the radiation field in the positive (left panel) and negative (right panel) direction for the 1D radiative transfer model discussed in Sect.~{\ref{simplecase.sec}}. All models have the same scattering cross section $C_{\text{sca}}=1$ and the different lines correspond to models with different absorption cross sections $C_{\text{abs}}$, ranging from 2 to --2.5 in steps of --0.25. The thick orange and red curves correspond to pure scattering and critical net stimulated emission, respectively.}
\label{variableCabs.fig}
\end{figure*}

For any value of the cross sections $C_{\text{abs}}$ and $C_{\text{sca}}$, this set of coupled equations can be solved analytically. We note that, as discussed in the previous sections, the scattering cross section is always positive, whereas the absorption cross section can also be negative, as it combines the contribution of actual absorption and stimulated emission. We assume that $C_{\text{sca}}>0$ and we consider five different cases, depending on the value of $C_{\text{abs}}$: net absorption ($C_{\text{abs}}>0$), pure scattering ($C_{\text{abs}}=0$), weak net stimulated emission ($-C_{\text{sca}}<C_{\text{abs}}<0$), critical net stimulated emission ($-C_{\text{sca}}=C_{\text{abs}}<0$), and strong net stimulated emission ($C_{\text{abs}}<-C_{\text{sca}}<0$). These five different regimes are indicated on Fig.~{\ref{variableCabs-paramspace.fig}}, which shows the $(C_{\text{abs}},C_{\text{sca}})$ parameter space for this simple radiative transfer problem.

\subsubsection{Net absorption}

In the regime of net absorption, thus with $C_{\text{abs}}>0$, the set of coupled differential equations~(\ref{RTE1D}) has two real eigenvalues. Given the boundary conditions (\ref{RTE1D-bound1}) and (\ref{RTE1D-bound2}), we can write the full solution as
\begin{gather}
I^+(x) = \frac{\xi \cosh \xi (1-x) + \zeta \sinh \xi (1-x)}{\xi \cosh \xi  + \zeta \sinh \xi },
\label{Iplus}
\\[0.5em]
I^-(x) = \frac{\tfrac12\,C_{\text{sca}} \sinh \xi (1-x)}{\xi \cosh \xi + \zeta \sinh \xi },
\label{Imin}
\end{gather}
where we have defined the constants
\begin{gather}
\xi = \sqrt{C_{\text{abs}}\,(C_{\text{abs}}+C_{\text{sca}})}, \\
\zeta = C_{\text{abs}} + \tfrac12\,C_{\text{sca}}.
\label{zeta}
\end{gather}
Both these constants are strictly positive, and so are the solutions~(\ref{Iplus}) and (\ref{Imin}) for all values of $C_{\text{abs}}$ and $C_{\text{sca}}$ in this regime.

\subsubsection{Pure scattering}

When $C_{\text{abs}}=0$, the differential equations~(\ref{RTE1D}) simplify to 
\begin{equation}
\frac{\txd I^\pm}{\txd x}(x) = 
-\tfrac12\,C_{\text{sca}}\,\left[I^+(x)-I^-(x)\right].
\end{equation}
Subject to the boundary conditions (\ref{RTE1D-bound1}) and (\ref{RTE1D-bound2}), the solution becomes 
\begin{gather}
I^+(x) = \frac{2 + C_{\text{sca}}\,(1-x)}{2+C_{\text{sca}}}, \label{Ipluspurescatt}\\
I^-(x) = \frac{C_{\text{sca}}\,(1-x)}{2+C_{\text{sca}}}. \label{Iminpurescatt}
\end{gather}
The same solution can also be found by taking the limit $C_{\text{abs}}\to0$ in the expressions~(\ref{Iplus}) -- (\ref{zeta}). It is immediately clear that this solution is positive, and thus physically valid, for all values of $C_{\text{sca}}$.

\subsubsection{Weak net stimulated emission}

In the weak net stimulated emission regime, we can still use the expressions~(\ref{Iplus}) -- (\ref{zeta}) as the formal solution of the coupled set of radiative transfer equations~(\ref{RTE1D}). However, since $C_{\text{abs}}<0$ and $C_{\text{abs}}+C_{\text{sca}}>0$, $\xi$ is an imaginary number. A more elegant way to write the solution is 
\begin{gather}
I^+(x) = \frac{\xi' \cos \xi' (1-x) + \zeta \sin \xi' (1-x)}{\xi' \cos \xi' + \zeta \sin \xi'},
\label{Ipluswnsa}
\\[0.5em]
I^-(x) = \frac{\tfrac12\,C_{\text{sca}} \sin \xi' (1-x)}{\xi' \cos \xi' + \zeta \sin \xi'},
\label{Iminwnsa}
\end{gather}
where we have set
\begin{equation}
\xi' = \sqrt{-C_{\text{abs}}\,(C_{\text{abs}}+C_{\text{sca}})} > 0.
\end{equation}
Interestingly, instead of the standard exponential functions usually encountered in radiative transfer problems, we now have trigonometric functions, as typically encountered in solutions of coupled differential equations with imaginary eigenvalues. 

Furthermore, it turns out that this formal solution does not always correspond to a physically sound solution: only when both $I^+(x)$ and $I^-(x)$ are positive at all positions $x$ along the column, the formal solution can be physical. This limits the region in the $(C_{\text{abs}},C_{\text{sca}})$ parameter space that corresponds to physical solutions. For $C_{\text{sca}}\leqslant2$, the solutions are always physical in the weak net stimulated emission regime. For every value of $C_{\text{sca}}>2$, however, there is a limit value of $C_{\text{abs}}$ below which the solutions (\ref{Ipluswnsa}) and (\ref{Iminwnsa}) become negative. This limit value is reached when the denominator of these expressions becomes zero, that is when 
\begin{equation}
\frac{\tan\xi'}{\xi'} = -\frac{1}{\zeta}.
\label{boundary}
\end{equation}
For each $C_{\text{sca}}>2$, this equation can be solved numerically to determine the limit value of $C_{\text{abs}}$. The boundary between the physical and non-physical models in the weak net stimulated emission regime is indicated as the upper part of the thick green line in Fig.~{\ref{variableCabs-paramspace.fig}}.

\subsubsection{Critical net stimulated emission}

The critical net stimulated emission case corresponds to the situation where absorption, scattering and stimulated emission are perfectly in balance such that their combined cross section is zero, that is, $C_{\text{abs}}+C_{\text{sca}} = 0$. For this special case, we can rewrite the differential equations~(\ref{RTE1D}) as
\begin{equation}
\frac{\txd I^\pm}{\txd x}(x) = 
\pm\tfrac12\,C_{\text{sca}}\,\left[I^+(x)+I^-(x)\right].
\end{equation}
Taking into account the boundary conditions~(\ref{RTE1D-bound1}) and (\ref{RTE1D-bound2}), the solution is
\begin{gather}
I^+(x) = \frac{2-C_{\text{sca}}\,(1-x)}{2-C_{\text{sca}}},
\label{Ipluscrit}
\\[0.5em]
I^-(x) = \frac{C_{\text{sca}}\,(1-x)}{2-C_{\text{sca}}}.
\end{gather}
Also in this case, the solutions are not physically valid for all values of $C_{\text{sca}}$. In fact, it is easy to see that they are only physical when $C_{\text{sca}}<2$, or equivalently, when $C_{\text{abs}}>-2$.

\subsubsection{Strong net stimulated emission}

Finally, we consider the strong net stimulated emission regime where the net stimulated emission cross section is so large that it dominates over the combined absorption and scattering cross sections. In this case, the solution of the radiative transfer problem is formally exactly the same as in the net absorption regime, that is, the expressions~(\ref{Iplus}) and (\ref{Imin}). 

Since both $C_{\text{abs}}$ and $C_{\text{abs}} + C_{\text{sca}}$ are negative in this regime, we have that $\xi>0$. The main difference with the case of net absorption is that $\zeta$ is negative in this case, which implies that the denominator will become negative for certain combinations of $C_{\text{abs}}$ and $C_{\text{sca}}$. For every value of $C_{\text{sca}}<2$, there is a limit value of $C_{\text{abs}}$ below which the solutions (\ref{Iplus}) and (\ref{Imin}) become negative. This limit value is reached when 
\begin{equation}
\frac{\tanh\xi}{\xi} = -\frac{1}{\zeta}.
\label{boundary2}
\end{equation}
For $C_{\text{sca}}>2$, the radiative transfer problem has no physical solutions in the strong net stimulated emission regime. The boundary between physical and non-physical solutions in the strong net stimulated emission regime smoothly connects to the similar boundary in the weak net stimulated emission regime, and is indicated as the thick green line in Fig.~{\ref{variableCabs-paramspace.fig}}.

\subsubsection{No scattering}

To streamline the discussion of the various regimes in the previous subsections, we assumed that $C_{\text{sca}}>0$. We now consider the case where there is no scattering, i.e., $C_{\text{sca}}=0$. The radiative transfer equations (\ref{RTE1D}) simplify drastically and are no longer coupled. Taking into account the boundary conditions~(\ref{RTE1D-bound1}) and (\ref{RTE1D-bound2}), the trivial solution is
\begin{gather}
I^+(x) = \exp\,(-C_{\text{abs}}\,x),
\label{Ipluscrit}
\\[0.5em]
I^-(x) = 0.
\end{gather}
This solution is physically valid for all finite values of $C_{\text{abs}}$. The pure scattering, weak net stimulated emission, and critical net stimulated emission regimes collapse into the single point where $C_{\text{sca}}=C_{\text{abs}}=0$. This no extinction regime is represented in Fig.~{\ref{variableCabs-paramspace.fig}} by the intersection of the orange and red lines.

\begin{figure*}
\includegraphics[width=\textwidth]{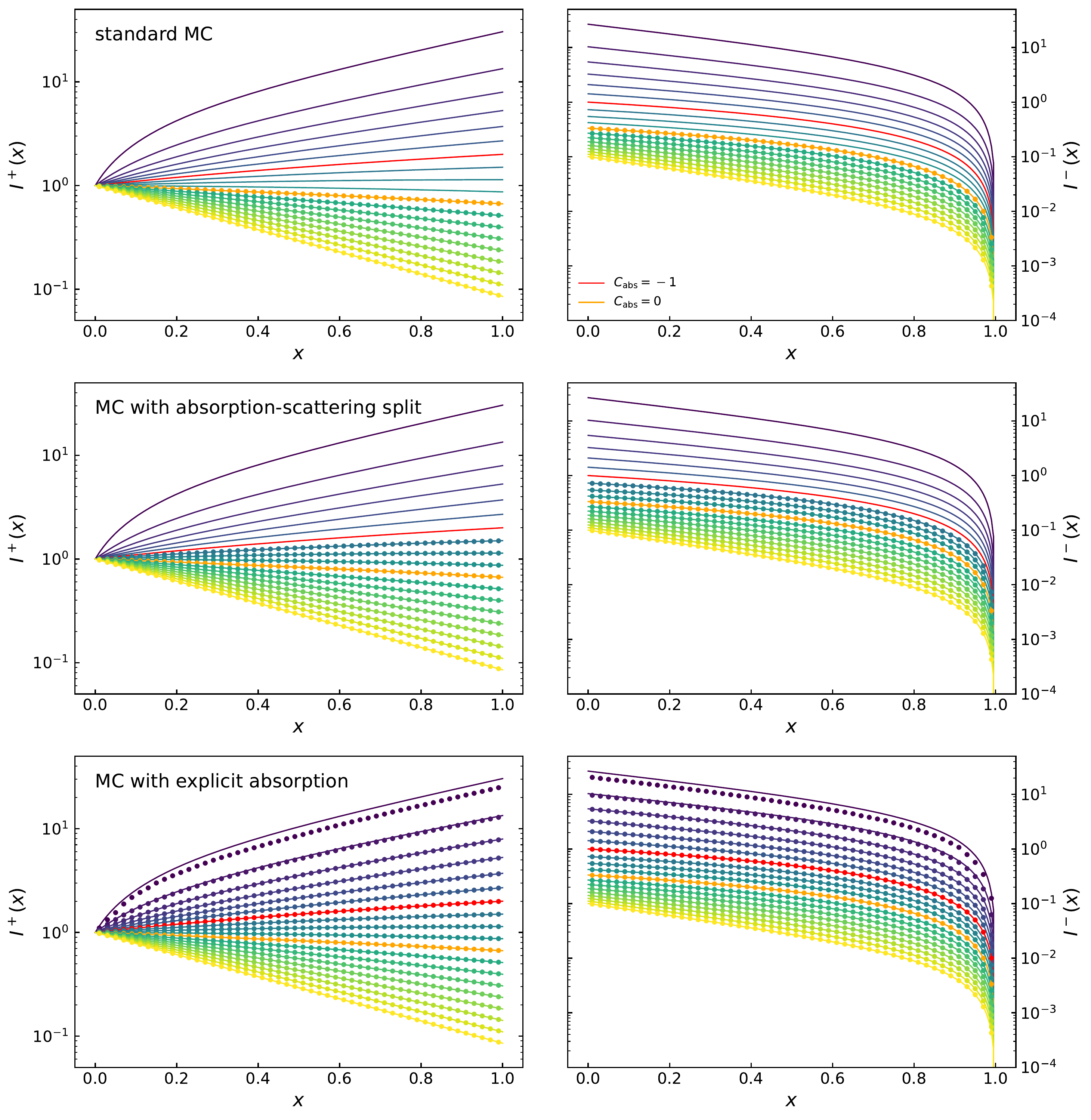}
\caption{Monte Carlo simulation results of the 1D radiative transfer problem discussed in Sect.~{\ref{simplecase.sec}}. The three rows correspond to three different variants of the MCRT method: the standard implementation (top row), the version with the absorption-scattering split (middle row), and the version with explicit absorption (bottom row). Solid lines are the analytical results, the dots are the results of the MCRT calculations. }
\label{MCRT.fig}
\end{figure*}

\subsection{Qualitative discussion of the solution}
\label{simplecase.sec}

In Fig.~{\ref{variableCabs.fig}} we illustrate the systematic change of the solution of the two-stream radiative transfer problem as we vary the model parameters. We have fixed $C_{\text{sca}}=1$, and the different lines in this figure represent solutions of the RTE when we gradually decrease the absorption cross section $C_{\text{abs}}$ from 2 to --2.5 in steps of $\Delta C_{\text{abs}} = -0.25$. In Fig.~{\ref{variableCabs-paramspace.fig}} we show the position of each of these models in the $(C_{\text{abs}}, C_{\text{sca}})$ parameter space. This set of parameters covers each of the five regimes discussed in the previous subsection.

For positive values of $C_{\text{abs}}$ we are obviously in the net absorption regime. The largest value of $C_{\text{abs}}$ corresponds to the strongest attenuation of the intensity in the positive direction. The intensity profile $I^+(x)$ is almost a perfect exponentially decaying function, as expected when absorption strongly dominates over scattering. The intensity in the negative direction, which is entirely due to scattered radiation, is also a monotonically decreasing function of $x$. When $C_{\text{abs}}$ decreases, the total extinction cross section decreases as well. The result is that $I^+(x)$ gradually becomes flatter and less exponential-like, and that $I^-(x)$ gradually increases with decreasing $C_{\text{abs}}$. 

When $C_{\text{abs}}$ becomes zero, the model contains only scattering. Compared to the models with net absorption, the intensity profile in the positive direction is even flatter. As shown by expression~(\ref{Ipluspurescatt}), the $I^+(x)$ profile is now a monotonically decreasing linear function of $x$ rather than an exponential function. The $I^-(x)$ profile is also a linearly decreasing function of $x$ that is globally increased compared to the solutions of the net absorption regime. 

As soon as $C_{\text{abs}}$ decreases even more and hence becomes negative, we enter the net stimulated emission regime. We first enter the weak net stimulated emission regime, in which the scattering cross section dominates over the negative absorption cross section. As $C_{\text{abs}}$ gradually becomes more negative, the intensity profile in the forward direction becomes flatter and flatter, as a result of the competition between stimulated emission and scattering. At a certain point, $I^+(x)$ becomes a non-monotonic function of $x$: it first slightly increases as a function of $x$ and subsequently slightly decreases. More specifically, this starts at $C_{\text{abs}} = -0.2824$. A nice illustration of such a model is the case $C_{\text{abs}} = -0.3799$: for this specific value, we find that $I^+(1) = 1$, that is, the forward intensity at the start and the end point of the column has exactly the same value.  It does reach a maximum value of 1.0302 at $x=0.5$. If $C_{\text{abs}}$ becomes even more negative, there is no longer attenuation and the intensity at the end point of the column exceeds the input intensity. As soon as $C_{\text{abs}}<-0.5$, $I^+(x)$ becomes a monotonically increasing function of $x$, with a strong initial increase and a gradual flattening at large values of $x$. Over the entire range of $C_{\text{abs}}$ values, the negative intensity profile remains a decreasing function of $x$, and the magnitude steadily increases as $C_{\text{abs}}$ becomes more negative.

When $C_{\text{abs}}=-1$, we end up in the situation where the absorption, scattering and stimulated emission cross sections are perfectly in equilibrium, that is, we are in the critical net stimulated emission regime. As can be seen from Eq.~(\ref{Ipluscrit}), $I^+(x)$ is now a monotonically increasing linear function of $x$. The intensity in the negative direction is a monotonically decreasing linear function of $x$.

Finally, when $C_{\text{abs}}$ gets even more negative, we enter the strong net stimulated emission regime. The smaller $C_{\text{abs}}$, the stronger the forward intensity profile $I^+(x)$ will grow as a function of $x$. It is interesting to see that the slope of the $I^+(x)$ curve is strongest at small $x$ and gradually flattens at larger $x$. The intensity profile in the negative direction remains a decreasing function of $x$, but its magnitude increases steadily as $C_{\text{abs}}$ grows more negative. At some point, the exponential growth rate of $I^+(x)$ at small $x$ is so strong that the intensity diverges before it reaches the other side of the column. This is the point where we reach the green curve in Fig.~{\ref{variableCabs-paramspace.fig}}, corresponding to the relation~(\ref{boundary2}). This is where the solution (\ref{Iplus})--(\ref{Imin}) becomes unphysical. For the current choice of parameters, the critical boundary between physical and unphysical intensity profiles is $C_{\text{abs}}=-2.7340$.

\subsection{Monte Carlo simulations}

For the simple 1D radiative transfer problem defined in the previous section, we constructed a simple Monte Carlo code. We implemented three different variants: one version with the standard implementation as described in Sec.~{\ref{BasicMC.sec}}, one version with the absorption-scattering split (Sec.~{\ref{SplitMC.sec}}), and one version with explicit absorption (Sec.~{\ref{ExplicitMC.sec}}). In each case, the radiation field is recorded at 50 locations spread uniformly along the column.

In Fig.~{\ref{MCRT.fig}} we show the results of our simulations. In each panel, the solid lines are the analytical results also shown in Fig.~{\ref{variableCabs.fig}} and the dots are the results of the MCRT calculations. The top panels show the results for the standard implementation. This procedure only works for $C_{\text{abs}}\geqslant 0$ and it clearly reproduces the analytical results perfectly. 

The panels on the middle row show the results of MCRT simulations including the absorption-scattering split. For the models with $C_{\text{abs}}\geqslant 0$, the analytical results are still reproduced accurately. As argued in Sec.~{\ref{SplitMC.sec}}, the introduction of the absorption-scattering split also allows us to extend the applicability of the MCRT method to the regime of weak net stimulated emission. We see that, indeed, the MCRT results also reproduce the analytical results for all values $C_{\text{abs}}>-C_{\text{sca}}=-1$.

Finally, the bottom panels show the results for the MCRT simulations with explicit absorption. In this case, the MCRT simulations can be applied for all parameter settings, including the regimes of critical and strong net stimulated emission. The MCRT results reproduce the analytical results accurately over the entire range of $C_{\text{abs}}$ values. Only for the most negative values of $C_{\text{abs}}$, we see deviations from the analytical result. The deviation is particularly notable for $C_{\text{abs}}=-2.5$, the most negative value considered. This value is close to the limit value $-2.7340$ that $C_{\text{abs}}$ can have before the problem diverges and becomes unphysical. 

%%%%%%%%%%%%%%%%%%%%%%%%%%%%%%%%%%%%%%%%%%%%%%%%%%%%%%%%%%%%%%%

\section{Implementation in SKIRT}
\label{sec:ImplementationSKIRT}

Having demonstrated the power of the explicit absorption technique using a simple 1D problem, we now describe its implementation in a state-of-the-art 3D code.

\subsection{The SKIRT code}

SKIRT\footnote{The open-source SKIRT code is registered at the ASCL with the code entry ascl:1109.003 and is hosted at \url{www.github.com/SKIRT/SKIRT9}. Documentation and other information can be found at \url{www.skirt.ugent.be}.} \citep{2011ApJS..196...22B, 2015A&C.....9...20C, 2020A&C....3100381C} is a feature-rich 3D MCRT code that is commonly being applied to construct synthetic observations of astrophysical models. Recent examples include \citet{2021MNRAS.503.2349D, 2021MNRAS.506.3946D, 2021MNRAS.503..511G, 2021A&A...653A..34V, 2021MNRAS.506.5703K, 2021MNRAS.507.2755L, 2021ApJ...919..139W, 2021MNRAS.507.5246M, 2021ApJ...922...88O, 2022ApJ...926..192V, 2022MNRAS.510.3321P, 2022MNRAS.510.5560S, 2022MNRAS.511.4999V, 2022MNRAS.512..245F}. SKIRT offers an extensive library of built-in geometries \citep{2015A&C....12...33B} as well as the capability of importing models from hydro-dynamical simulation snapshots \citep{2015A&A...576A..31S, 2016MNRAS.462.1057C}. Sources can be assigned spectra from built-in or user-provided stellar or stellar population template libraries \citep{2015A&C.....9...20C}. Media types and physical processes include dust extinction and emission using turn-key or highly configurable dust models \citep{2015A&A...580A..87C}, polarisation by dust grains \citep{2017A&A...601A..92P, 2020AJ....160...55V}, Lyman-$\alpha$ resonant line transfer \citep{2021ApJ...916...39C}, electron Thomson and Compton scattering, photo-ionisation and fluorescence by neutral atoms (Vander Meulen et al., in prep), and kinematics \citep{2003MNRAS.343.1081B, 2020A&C....3100381C}. Supported wavelengths span the X-ray, UV to sub-millimetre, and radio regimes.

On the numerical side, SKIRT offers several spatial discretisation options, including 1D spherical and 2D cylindrical grids for models with the corresponding symmetries, and advanced hierarchical and unstructured grids for high-dynamic-range 3D models \citep{2013A&A...554A..10S, 2014A&A...561A..77S, 2013A&A...560A..35C}. The configuration for a given model can include any mixture of source and media types. The Monte Carlo PLC supports primary emission by the sources, absorption and scattering by the media, and secondary emission by eligible media, while tracking the radiation field and recording any requested observables. Self-consistent iteration can be enabled to properly handle processes such as dust destruction near high-intensity sources (during primary emission) or dust self-absorption in regions with high optical depth (during secondary emission). The PLC runs in parallel on multiple execution threads and/or in multiple processes, allowing the code to efficiently scale from laptops to multi-node high-performance clusters \citep{2017A&C....20...16V}.

SKIRT uses several acceleration and variance-reduction techniques. These include peel-off towards the instruments at emission and scattering events for recording observables \citep{1984ApJ...278..186Y}, the absorption-scattering split described in Sect.~\ref{SplitMC.sec} \citep{1970A&A.....9...53M, 1977ApJS...35....1W}, continuous absorption along the photon packet path to improve sampling of the radiation field \citep{1999A&A...344..282L,2003A&A...399..703N}, forced scattering to improve sampling of optically thin regions \citep{Cashwell1959}, path length biasing to more easily penetrate optically thick regions \citep{2016A&A...590A..55B}, and wavelength biasing to improve signal-to-noise in wavelength ranges of interest. Many of these options can be configured by the user. For example, forced scattering can be turned off to speed up resonant line transfer simulations, where each photon packet undergoes a potentially large number of scattering interactions with very short free path lengths (often within the same grid cell). In addition to all this, the code switches to optimised versions of some procedures depending on the configuration. For example, if a photon packet's wavelength cannot change along its path because the model has no kinematics and optical media properties are constant across space, the absorption and scattering cross sections can be calculated once at the start of the path instead of repeating the calculation for each crossed cell.

\subsection{Adding explicit absorption}
\label{sec:addingexpabstoSKIRT}

In view of the complexity described in the previous subsection, adding the explicit absorption technique to the SKIRT PLC seems daunting. The most important question is whether explicit absorption can co-exist with the other acceleration and variance reduction techniques, most notably peel-off, continuous absorption, forced scattering, and path length biasing. A secondary concern is whether explicit absorption can be added as a user-configurable run-time option without impacting performance of the existing PLCs, while limiting the amount of code duplication. It turns out that, with the proper implementation choices, co-existence with the other techniques is not a problem and supporting the various types of PLCs in a single code is quite feasible.

In the updated SKIRT version\footnote{Explicit absorption was added to SKIRT 9 by git commit 48e8404 on June 9, 2022}, we allow enabling or disabling forced scattering and/or explicit absorption independently of each other. This yields four PLC variants, each of which has specific advantages or drawbacks.

Forced scattering tends to reduce noise for simulations with low optical depths. However, for each scattering event, it requires the calculation of the geometry and optical depth of the full path up to the model boundary, regardless of the location of the scattering event along the path. In models with very intensive scattering, such as for Lyman-$\alpha$ line transfer, the PLC without forced scattering is often the better choice, because it avoids calculating the path geometry and optical depth beyond the scattering location. However, our implementation does not support storing the radiation field. The main reason for this choice is that some additional optimisations can be applied. For example, there is no need to store the calculated path segment information. As a consequence, forced scattering cannot be turned off for simulations that require the radiation field to calculate secondary emission by media. 

The PLC without explicit absorption uses the extinction (sum of scattering and absorption) along a photon packet's path to locate the next interaction point. This requires the cumulative extinction optical depth to be a non-decreasing function of path length. It is thus not possible to handle negative extinction cross sections. The PLC with explicit absorption instead uses the scattering optical depth to locate the next interaction point. While the scattering cross section still must be non-negative, this allows the extinction cross section to be negative. As a drawback, this technique requires calculating both the scattering and absorption optical depths for the photon packet path.

\subsection{The photon packet}
\label{sec:SKIRTphotonpacket}

At the heart of the Monte Carlo PLC in SKIRT is the C++ class used for representing a photon packet. A photon packet tracks all relevant properties during its life cycle, including wavelength, weight (i.e., number of photons), polarisation state, position, and propagation direction. It can also store details on the path that traverses the simulation's spatial grid in the packet's current propagation direction and starting at its most recent interaction or emission location. Specifically, for each path segment $n$ crossing a given spatial grid cell, this information includes the cumulative distance $s_n$ and the cumulative optical depth $\tau_n$ up to the point where the path exits the cell. Once calculated and stored, these path segment properties can be used to determine the next interaction location and to add the packet's contribution to the radiation field in each crossed cell.

To support explicit absorption, we adjusted the path segment data structure so that it can now hold two distinct cumulative optical depth values, $\tau_n^\mathrm{ext|sca}$ and $\tau_n^\mathrm{abs}$. When explicit absorption is enabled, the PLC code calculates and stores the scattering and absorption optical depths $\tau_n^\mathrm{sca}$ and $\tau_n^\mathrm{abs}$ in these fields. When explicit absorption is not used, the code instead calculates and stores the extinction optical depth $\tau_n^\mathrm{ext}$ in the first field, and the second field is set to zero. As a result of this scheme, the procedure for locating the next interaction point can always use the $\tau_n^\mathrm{ext|sca}$ field, which will store the proper value depending on whether explicit absorption is enabled or not, and the procedure for adjusting the radiation field can always obtain the extinction optical depth through a simple addition, i.e. $\tau_n^\mathrm{ext|sca} + \tau_n^\mathrm{abs}$. With this `trick', we were able to substantially limit the changes to the code in areas that are not specifically concerned with the explicit absorption technique.

\subsection{The photon packet life cycle}

The top-level function driving the SKIRT PLC already had two separate code branches, one with and one without forced scattering. To allow selecting explicit absorption at run-time as well, we added an extra code path at just two places in each branch. When calculating the optical depths for the segments along the current path, we invoke a different function to ensure that the appropriate type of optical depth values are stored in the photon packet (see Sect.~\ref{sec:SKIRTphotonpacket}). And when applying the bias factor to the packet's weight for a scattering interaction, we either use the scattering albedo (without explicit absorption) or the absorption extinction up to the interaction location according to Eq.~(\ref{ws}) in Sect.~\ref{ExplicitMC.sec} (with explicit absorption). The required if-then-else statements are performed only once per photon packet path segment and thus do not affect performance. 

For most simulation configurations, a substantial fraction of the execution time is spent in the function calculating the path segments and the corresponding optical depths for the photon packet path. It is therefore justified to optimise this portion of the code for speed. There now is a separate version of this function for each of the four variants of the PLC. Furthermore, each of these functions have separate code paths depending on the overall configuration of the simulation. For example, there is a version optimised for configurations that allow calculating the absorption and scattering cross sections just once at the start of each path. In total, these various implementations occupy under 600 lines of code in a single class, including comments and documentation, resulting in an acceptable trade-off between complexity and performance.

No other changes to the code were needed, as we will verify in the next section. When explicit absorption is enabled, the photon packets follow different random paths because the interaction points are determined based on the scattering optical depth rather than the extinction optical depth. However, the photon weight is adjusted differently to compensate, so that the other aspects of the simulation continue to work properly. Specifically, the radiation field is properly tracked and the peel-off photon packets properly sample the fluxes observed by the configured instruments. Path length stretching, when enabled, also continues working as expected in the adjusted context.

%%%%%%%%%%%%%%%%%%%%%%%%%%%%%%%%%%%%%%%%%%%%%%%%%%%%%%%%%%%%%%%

\begin{figure*}
\centering
\includegraphics[width=0.95\textwidth]{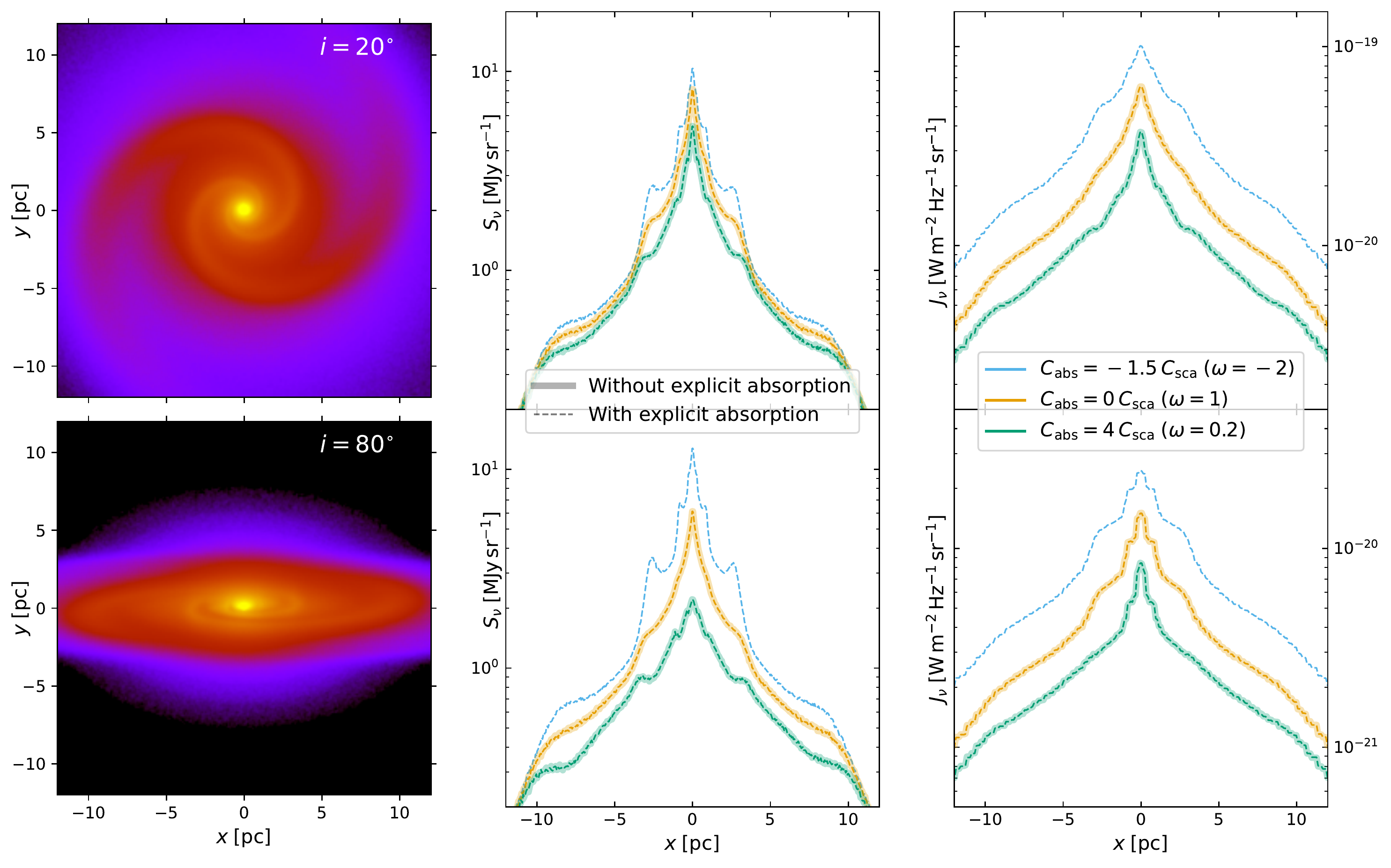}
\caption{SKIRT results for a smooth spiral galaxy model with varying optical medium properties, viewed at inclinations of $20^\circ$ (top row) and $80^\circ$ (bottom row). The left column shows the surface brightness assuming a dust medium at V-band wavelengths, using an arbitrary colour scale. The middle column shows the average surface brightness along a central horizontal slice, assuming the model is observed at a distance of 10~Mpc. Results are shown for three types of media replacing the dust (green, orange and blue colours) and calculated without (solid curves) or with (dashed curves) explicit absorption. See Sect.~\ref{sec:spiral} for details. The right column shows the mean intensity of the radiation field inside each of these model variations, averaged along the line of sight (weighted by medium mass) and averaged along a horizontal slice of the view.}
\label{Spiral.fig}
\end{figure*}

\section{Verification of the SKIRT implementation}
\label{sec:Application3D}

\subsection{Regression tests}

Over the last decade of SKIRT development, we continuously maintained and expanded our set of functional test cases used for regression testing. We now have about 700 test cases that cover most features and code paths and can be run in ten minutes or so on a present-day desktop computer. Each test simulation runs on a single execution thread and uses a fixed pseudo-random number sequence. This ensures that output files are binary identical between runs, enabling automated verification. Using this test suite, we easily confirmed that the updated implementation described in Sect.~\ref{sec:ImplementationSKIRT} does not break the already present PLCs.

\subsection{The 1D two-stream problem}

As an initial test for the explicit absorption technique, we set up a 3D SKIRT model that is equivalent to the 1D two-stream problem discussed in Sect.~\ref{sec:Application1D}. The model includes a narrow tube of material along the x-axis and a point source injecting a laser beam into the tube at one end. The material has configurable absorption and scattering cross sections at the laser's wavelength. When a photon packet scatters, the interaction reverses the packet's direction or leaves it unchanged with fifty per cent probability for each action. The simulation is further configured to record the radiation field along the tube in regularly spaced grid cells. This automatically enables forced scattering (see Sect.~\ref{sec:addingexpabstoSKIRT}). Because SKIRT does not record directional information for the radiation field, the resulting quantity is equivalent to the sum of the intensities, $I^+(x)+I^-(x)$, in the 1D problem.

Running this model for a set of relevant cross section values sampling the shaded area in Fig.~\ref{variableCabs-paramspace.fig}, we confirmed that SKIRT properly reproduces the analytic solutions listed in Sect.~\ref{sec:Application1D}. We note that the model used here can be constructed with standard SKIRT components except for the special material properties, which we defined through a trivial custom class. Because of its 1D scattering behaviour, this material definition is not very useful for general 3D modelling. A more useful variation described in the next section has been included as a standard SKIRT component.

\subsection{A spiral galaxy model}
\label{sec:spiral}

To test our implementation with an actual 3D geometry for which the results are still easily interpreted, we use the smooth spiral galaxy model presented by \citet{2017A&A...601A..92P} in their Sect.~6 with properties listed in their Table~2. In short, the model includes a stellar bulge and disk with an older star population, a second stellar disk with a younger star population, and a dust disk. The density distributions of the stellar and dust disks are perturbed by a smooth spiral arm structure. The left panels of Fig.~\ref{Spiral.fig} illustrate the observed V-band surface brightness as calculated by SKIRT for near face-on and near edge-on viewing angles.

We replace the dust in the model by an idealised medium for which we can directly configure the optical properties at the single, arbitrary wavelength used in our tests. We configure a slight preference for forward scattering ($\left<\cos\theta\right>=0.25$), set the scattering cross section $C_\text{sca}$ to a fixed value, and let the absorption cross section $C_\text{abs}$ vary as a multiple of $C_\text{sca}$, i.e. $C_\text{abs} = k\,C_\text{sca}$. The total mass of the medium is normalised such that the extinction optical depth for a configuration with $k=1$ is similar to that for the model with dust. As we vary $k$, the extinction optical depth varies as well.

The panels in the middle column of Fig.~\ref{Spiral.fig} show the average surface brightness calculated by SKIRT along a central horizontal slice for each of the two viewing angles. Results are shown for optical properties in three regimes, respectively with dominating absorption (green, $k=4$), no absorption (orange, $k=0$), and net stimulated emission (blue, $k=-1.5$). As expected, the surface brightness increases as $k$ decreases, and the features of the curves trace the spiral arms in the model. These effects are more prominent for the near edge-on view because this line of sight has longer paths through the medium. For the first two regimes (green and orange), the calculations were performed without and with explicit absorption, shown as solid and dashed curves, respectively. It is apparent from the figure that the results are essentially identical. The regime with net stimulated emission (blue) requires explicit absorption to be enabled, so only this result can be shown. Even if no direct comparison is possible, the progression seen in the curves for the three regimes does offer comfort that this latter result is correct as well.

The right panels of Fig.~\ref{Spiral.fig} show the mean intensity of the radiation field inside each of these model variations, averaged along the line of sight and along a horizontal slice of the view. These results confirm that the implementation with explicit absorption also properly records the radiation field.

All of the results shown in Fig.~\ref{Spiral.fig} were calculated with forced scattering enabled because of the need for recording the radiation field during the simulation. We separately confirmed, however, that the PLC without forced scattering also reproduces the surface brightness results of the middle panel.

\subsection{Published benchmarks}

In previous work, we employed several benchmark problems that have been published in the literature to test SKIRT with dust media \citep{2020A&C....3100381C} and for Lyman-$\alpha$ line transfer \citep{2021ApJ...916...39C}. For the current work, to verify our updated implementation, we ran the benchmark problems listed below with explicit absorption enabled.
\begin{itemize}
  \item \citet{1997MNRAS.291..121I}: a spherically symmetric (1D) circumstellar dust shell with various dust density profiles and optical depths.
  \item \citet{2004A&A...417..793P}: an axisymmetric (2D) circumstellar dust disk with various edge-on optical depths.
  \item \citet{2009A&A...498..967P}: a disk geometry similar to that in the previous item, now including the effects of polarisation for anisotropic scattering by spherical dust grains. 
  \item \citet{2017A&A...603A.114G}: a uniform dust slab (3D) externally illuminated by a star, testing dust extinction, heating, and emission for various optical depths of the slab.
 \item \citet{2006ApJ...649...14D}: analytical approximation for the spectrum emerging from a static, uniform hydrogen sphere with a central point source emitting at the Lyman-$\alpha$ line centre.
 \item \citet{2006ApJ...645..792T}: numerical solutions for static, expanding and contracting hydrogen spheres with a central Lyman-$\alpha$ point source or with uniform emission throughout the sphere.
\end{itemize}
The dust benchmarks were run with forced scattering and the Lyman-$\alpha$ benchmarks without forced scattering. All results are within the uncertainty levels noted in previous work \citep{2020A&C....3100381C, 2021ApJ...916...39C}. Even if all materials in these benchmark models have non-negative cross sections, we can still conclude that the explicit absorption technique is very robust, at least in this regime.
 
%%%%%%%%%%%%%%%%%%%%%%%%%%%%%%%%%%%%%%%%%%%%%%%%%%%%%%%%%%%%%%%

\begin{figure*}
\centering
\includegraphics[width=\textwidth]{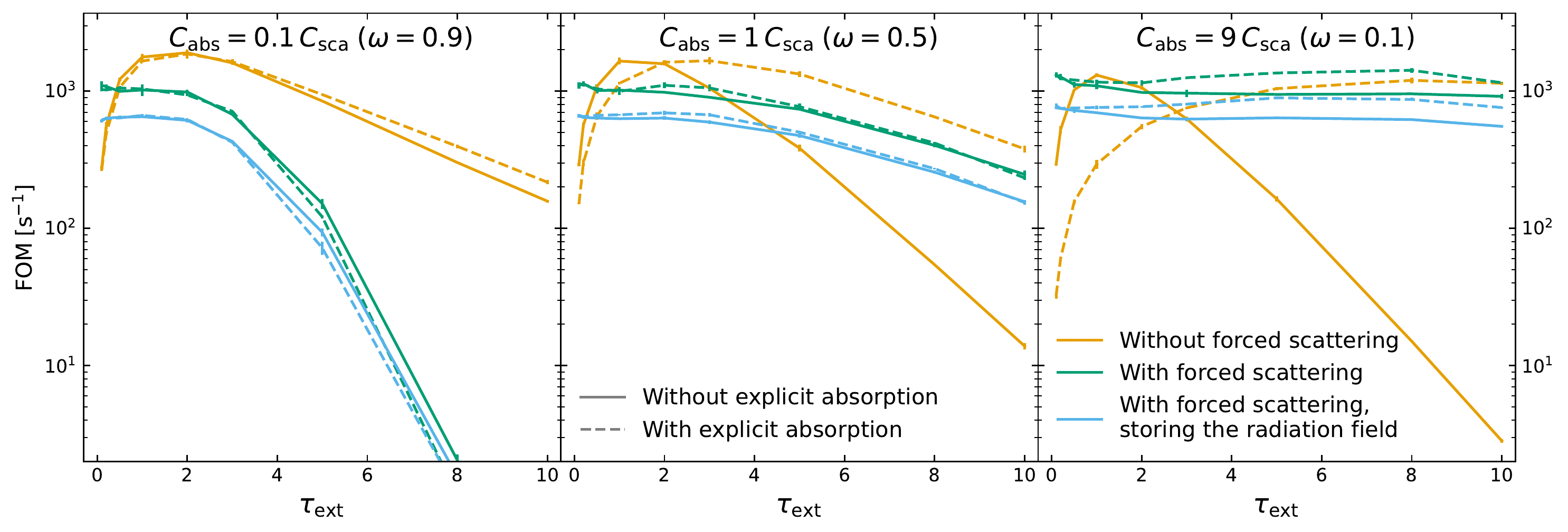}
\caption{Figure of merit (FOM, defined in Sect.~\ref{sec:FOM}) as a function of the transverse extinction optical depth of our slab model, $\tau_\text{ext}=0.1, 0.2, 0.5, 1, 2, 3, 5, 8, 10$.  The simulations were run on a basic multi-core desktop computer. A higher FOM value indicates a more efficient algorithm. Results are given for several variants of the photon packet life cycle (PLC): colours indicate without (orange) or with (green) forced scattering, or with forced scattering and also recording the radiation field (blue). The three panels correspond to different values of the scattering albedo. The left panel ($\omega=0.9$) represents the strong scattering regime and is applicable to dust attenuation at UV wavelengths. The central panel ($\omega=0.5$) corresponds to a situation where absorption and scattering are equally important and corresponds, for example, to dust attenuation at optical wavelengths. The right panel ($\omega=0.1$) corresponds to weak scattering, as applicable to dust attenuation at infrared wavelengths or radiative transfer in molecular and atomic lines. Line styles indicate without (solid) or with (dashed) explicit absorption. Error bars indicate the estimated uncertainty on the FOM values caused by the remaining Monte Carlo noise in the converged simulations.}
\label{FOM.fig}
\end{figure*}

\section{Efficiency}
\label{sec:Efficiency}

\subsection{The figure of merit (FOM)}
\label{sec:FOM}

We noted in Sect.~\ref{ExplicitMC.sec} that the explicit absorption technique should reduce Monte Carlo noise because it eliminates a stochastic element (the absorption/scattering split) but, on the other hand, requires the calculation of both the absorption and the scattering optical depth along the photon packet path. We now investigate how these differences influence the overall performance of the algorithm. To accomplish this, we need a mechanism to quantify and measure such performance.

\citet{2018ApJ...861...80C} discuss quantitative statistical tests for the convergence of MCRT results, borrowed from the manual for the Monte Carlo N-Particle transport code MCNP \citep{MCNP2003}. We refer to these works and references therein for details. The method entails tracking the moments $\sum_i w_i^k, k=1,2,3,4$ of the individual photon packet contributions $w_i$ to each result bin during the simulation. From these moments and the total number $N$ of photon packets launched, two dimensionless quantities are calculated for each bin: the relative error, $R$, and the variance of the variance, VOV. If both values are below 0.1, the result can be considered to be converged and $R$ can be interpreted as a true relative error on the result recorded for the bin. Moreover, one can then define the figure of merit,
\begin{equation}
\text{FOM} = \frac{1}{R^2\,T},
\end{equation}
where $T$ is the computer time spent on the simulation in seconds. Because the computer time is proportional to the number of photon packets $N$ and the square of the relative error should scale as $1/N$, the FOM is approximately constant once the solution has converged. The FOM thus provides a measure for the efficiency of a particular method for solving the problem at hand. Specifically, a higher FOM indicates that the simulation can reach a given level of signal-to-noise in less computer time.

SKIRT allows tracking the information underlying these statistics for the observed fluxes \citep{2020A&C....3100381C}. We construct a basic slab model similar to the geometry used by \citet{2017A&A...603A.114G} and \citet{2020A&C....3100381C} for benchmarks and tests. In our version, a monochromatic source illuminates the top of a uniform slab of material with configurable optical properties and we observe the radiation emerging from the slab on the other side. The instrument views a central slice of the slab and has 150 bins along its length (each bin covers the width of the slice). After each simulation, we calculate mean $\left<R\right>$ and $\left<\text{VOV}\right>$ statistics averaged over these bins. For a given configuration of material properties and PLC variant, we run simulations with a successively larger number of photon packets until $\left<R\right><0.1$ and $\left<\text{VOV}\right><0.1$. Then we run two extra simulations with three times more photon packets and use the results from these three converged simulations to obtain an average FOM value and an estimate of its uncertainty caused by remaining Monte Carlo noise.

The material in our slab has uniform density and optical properties. For the purpose of calculating the penetrating flux we could therefore limit the spatial discretisation to a single grid cell enclosing the complete slab. However, we instead configure a regular grid with 75 cells in each of the horizontal directions and 100 cells in the vertical direction (the direction towards the source). This ensures that the calculation of the path geometries and optical depths for a fair number of cell segments are taken into account in the efficiency measure. The grid also allows the radiation field to be recorded with a fair spatial resolution.

This brings us to an important caveat. Our efficiency measure does not take into account the noise levels in the recorded radiation field, because SKIRT is not able to track the required statistics. It is possible that a given Monte Carlo method produces high-quality observed fluxes but records a more noisy radiation field. This will not be captured by our FOM metric.

Because of the dependency on the simulation time $T$, FOM values cannot be compared between computer systems. FOM ratios, however, should be stable. On the other hand, different model setups (i.e. other than our slab) and different implementations of the PLC (i.e. in codes other than SKIRT) might change relative FOM values in unpredictable ways. In spite of this, we believe the overall trends noted in the following subsection to be fairly robust.

\subsection{Comparing algorithm variants}
\label{sec:comparing}

In Fig.~\ref{FOM.fig}, we show the FOM as a function of transverse extinction optical depth for the slab model described in the previous subsection. The figure shows three values of the scattering albedo (panels from left to right) and several variants of the PLC (colours and line styles). Similar to our earlier approach for the spiral galaxy model, we configure a fixed value for $C_\text{sca}$ and a multiple of this value for the absorption cross section, $C_\text{abs} = k\,C_\text{sca}$. The total mass of the medium is then normalised such that the extinction optical depth (taking into account both scattering and absorption) across the slab in the vertical direction equals the required value, ranging from $\tau_\text{ext}=0.1$ to $10$. We arbitrarily configure a fairly strong preference for forward scattering, with $\left<\cos\theta\right>=0.5$.

We first consider the previously implemented PLCs, i.e.\ those without explicit absorption (solid curves). As expected, forced scattering (solid green) substantially improves performance over the basic method (solid orange) in all albedo regimes for low optical depths ($\tau\lesssim 0.5$). For higher optical depths, the situation is much less clear. As long as absorption dominates over scattering (middle and right panels), the forced scattering technique can be considered the method of choice, except perhaps for a narrow but possibly important intermediate optical depth range ($0.5\lesssim\tau\lesssim 3$) where it is slower by up to a factor of two. However, when scattering dominates (left panel), forced scattering slows down the simulation by more than an order of magnitude for $\tau\gtrsim 5$. This can be attributed at least in part to the large number of peel-off photon packets that need to be sent in this situation. It is thus wise to disable forced scattering for media of this type. The third solid curve (blue) shows the performance of the forced scattering PLC when storing the radiation field. As can be expected, this additional task slows down the simulation by a constant factor.

More interesting in the context of this paper is the performance of the explicit absorption technique (dashed curves). Comparing the two PLCs without forced scattering (dashed and solid orange curves), we conclude that explicit absorption tends to slow down the simulation for lower optical depths ($\tau\lesssim 2$) and accelerate it for higher optical depths. The performance differences rise with decreasing scattering albedo, reaching an acceleration by three orders of magnitude at $\tau\approx 10$ when absorption dominates (right panel). This remarkable improvement seems to be a direct result of taking absorption out of the probabilistic part of the PLC and handling it separately in a deterministic, explicit manner. When forced scattering is enabled (dashed and solid green curves, or equivalently, dashed and solid blue curves), the performance differences caused by explicit absorption are much smaller. Even for a very low scattering albedo (right panel), the acceleration is limited to less than a factor of two.

When reviewing the performance of all PLC variants, one notable feature stands out. In the regime of balanced scattering and absorption (middle panel), enabling explicit absorption for the PLC without forced scattering (orange curves) boosts performance for $\tau\gtrsim 3$ beyond that of the PLC with forced scattering (green curves). A similar but somewhat weaker boost can be seen for a medium with low scattering albedo (right panel). 

Because the relative performance of the PLC variants so strongly depends on the regime imposed by the medium properties, there does not seem to be a single optimal algorithm. The specific model geometry and many other factors (see the caveats discussed at the end of Sect.~\ref{sec:FOM}) will also further influence the comparison. Nevertheless, based on Fig.~\ref{FOM.fig} and our analysis, we propose as an overall rule of thumb to enable forced scattering except for media that are strongly dominated by scattering and with optical depths $\tau\gtrsim 0.5$, and to always enable explicit absorption.

With this rule the simulation may perform sub-optimally (by a factor of less than two) in the balanced scattering/absorption regime (middle panel). To get optimal performance, one would have to configure different options depending on optical depth (the solid and dashed orange curves cross over at $\tau\approx 2$). And evidently, explicit absorption would always be required if the medium exhibits net stimulated emission. While a generic rule of thumb should avoid such complexity, it might in some cases be worthwhile to measure the FOM and select the appropriate algorithm, for example, when processing a large set of similar models.

%%%%%%%%%%%%%%%%%%%%%%%%%%%%%%%%%%%%%%%%%%%%%%%%%%%%%%%%%%%%%%%

\section{Summary and conclusions}
\label{sec:Conclusions}

Most 3D radiative transfer codes employ the probabilistic Monte Carlo method, which follows the flow of photon packets through the obscuring medium rather than attempting to directly solve the radiative transfer equation. In conjunction with appropriate variance reduction and acceleration techniques, the approach is very powerful and allows one to emulate a wide range of physics. However, the classic Monte Carlo radiative transfer (MCRT) photon packet life cycle (PLC) requires a non-negative extinction cross section of the medium at every location in the simulated domain. This prohibits the straightforward handling of media with net stimulated emission, where the energy emitted at a given wavelength dominates the energy absorbed at that wavelength, so that the effective cross section becomes negative.

In this paper, we presented the explicit absorption technique that allows an adjusted MCRT PLC to efficiently handle net stimulated emission. The new technique uses the scattering optical depth along a photon packet's path to randomly select the next interaction location, as opposed to using the extinction (absorption plus scattering) optical depth, and offers a separate, deterministic treatment of the absorption. While the scattering cross section still must be non-negative, a requirement that is easily met, the absorption and extinction cross sections are allowed to be negative. As a result, the method can handle net stimulated emission without any problems. 

After describing the explicit absorption technique in detail, we verified its correctness and applicability in several ways. We first formulated a simple 1D radiative transfer problem and provided analytic solutions in various regimes, including net absorption and net stimulated emission. We verified that a special-purpose MCRT code incorporating explicit absorption is indeed capable of recovering these solutions in all regimes (see Fig.~\ref{MCRT.fig}). We then reported on the implementation of explicit absorption in SKIRT, a fully featured state-of-the-art 3D MCRT code. This turned out to be a rather straightforward endeavour. Specifically, explicit absorption was easily combined with the variance reduction and acceleration techniques already incorporated in SKIRT, including peel-off and forced scattering. We verified the operation of the SKIRT implementation in the net absorption regime through regression tests and published benchmarks, and in the net stimulated regime through the same 1D problem mentioned earlier and using a smooth 3D spiral galaxy model (see Fig.~\ref{Spiral.fig}).

In the final section, we studied the efficiency of the explicit absorption technique using a basic setup for observing the radiation penetrating a 3D slab of material with varying optical properties (see Fig.~\ref{FOM.fig}). We found that explicit absorption slightly accelerates the simulation in most regimes as long as forced scattering is appropriately enabled or disabled. Based on these results, we recommended to always include explicit absorption, even if it is not strictly necessary for simulations in the net absorption regime (see the rule of thumb at the end of Sect.~\ref{sec:comparing}). Future experience with the technique will tell whether these trends hold for more realistic and complex models as well.

%%%%%%%%%%%%%%%%%%%%%%%%%%%%%%%%%%%%%%%%%%%%%%%%%%%%%%%%%%%%%%%

\section*{Acknowledgements}

KM is a PhD Fellow of the Flemish Fund for Scientific Research (FWO-Vlaanderen) and acknowledges the financial support provided through grant number 1169822N.

\bibliography{ExplicitAbsorption_bib}

\end{document}